\documentclass[aps,prl,twocolumn,showpacs,superscriptaddress,amssymb]{revtex4}
\usepackage{graphicx}
\usepackage{hyperref}
\usepackage{natbib}
\usepackage{bm}
\begin{document}

\title{Single-spin magnetometry with multi-pulse sensing sequences
}

\author{G. de Lange}
\affiliation{Kavli Institute of Nanoscience Delft, Delft University of Technology, 
P.O. Box 5046, 2600 GA Delft, The Netherlands}
\author{D. Rist\`e}
\affiliation{Kavli Institute of Nanoscience Delft, Delft University of Technology, 
P.O. Box 5046, 2600 GA Delft, The Netherlands}
\author{V. V. Dobrovitski}
\affiliation{Ames Laboratory US DOE, Iowa State University, Ames IA 50011, USA}
\author{R. Hanson}
\affiliation{Kavli Institute of Nanoscience Delft, Delft University of Technology, 
P.O. Box 5046, 2600 GA Delft, The Netherlands}

\date{\today}

\begin{abstract}
We experimentally demonstrate single-spin magnetometry with multi-pulse sensing sequences. The use of multi-pulse sequences can greatly increase the sensing time per measurement shot, resulting in enhanced ac magnetic field sensitivity. We theoretically derive and experimentally verify the optimal number of sensing cycles, for which the effects of decoherence and increased sensing time are balanced. We perform these experiments for oscillating magnetic fields with fixed phase as well as for fields with random phase. Finally, by varying the phase and frequency of the ac magnetic field, we measure the full frequency-filtering characteristics of different multi-pulse schemes and discuss their use in magnetometry applications.
\end{abstract}


\maketitle

The ability to sense weak magnetic fields with nanometer scale resolution has
important applications in fundamental and biomedical sciences as well as information storage technology. 
Several architectures for highly sensitive magnetometers have been implemented, such as superconducting quantum interference devices (SQUIDS)  \cite{SQUID}, Hall sensors \cite{Hall}, sensors based on magnetic resonance force microscopy (MRFM) \cite{MRFM} and atomic vapors \cite{Vapour}. Recently, approaches to magnetometry based on tracking the evolution of a single electron spin have been proposed \cite{chernobrod, degen, Taylor}. 

The magnetic field to be detected shifts the energy levels of the spin through the Zeeman effect. When a superposition of
spin states is prepared, the Zeeman shift leads to a phase difference proportional to the magnetic field, which can
be detected as a population difference after application of a suitable control pulse.

\begin{figure}
\label{fig:f1}
\includegraphics[width=3.4in]{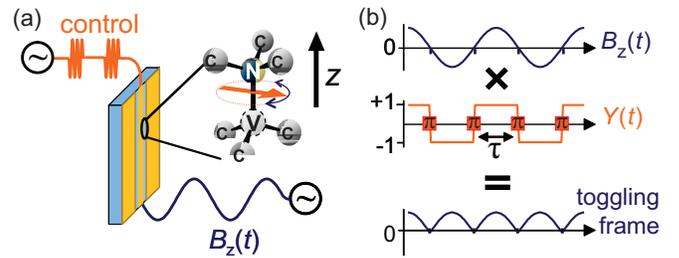}
\caption{\textbf{(a)} Magnetometry setup. A CPW transmission line, fabricated directly on a Ib diamond sample, is connected to control lines on both ends. One control line supplies 2.5 GHz microwave bursts for spin control. For details of this part of the setup see Ref.~\cite{bootstrap}. We use the other side of the CPW to supply the low frequency ($<5$ MHz) ac field $B_z(t)$ that we aim to detect. A single NV center located in between the conductors of the CPW is used as the magnetometer. \textbf{(b)} Multipulse magnetometry. 
An ac field $B_z(t)$ modulates the phase of the NV electron spin.
Timing $\pi$-pulses such that they coincide with the nodes of $B_z(t)$ effectively multiplies the field by $Y(t)$. The bottom trace shows the resulting $B_z(t)$ in the toggling reference frame, which flips whenever a $\pi$-pulse is applied to the spin.
}
\end{figure}

The nitrogen vacancy (NV) defect center in diamond is a prime candidate for single-spin magnetometry \cite{degen, Taylor,qmagnetNV,qmagnetNV2}. The NV center combines an excellent field sensitivity with high spatial resolution resulting from the near-atomic size of the defect. It has a ground state paramagnetic spin ($S$=1) with the $m_s=0$ and $m_s = \pm 1$ levels split by 2.87 GHz at zero magnetic field, with the quantization axis oriented along the symmetry axis of the defect. The NV center spin can be initialized and its spin population difference can be detected optically by measuring its spin-dependent photoluminescence (PL)~\cite{revNV}. First experimental demonstrations of NV-based magnetometry using a sensing sequence with a single $\pi$-pulse (spin echo) have outlined its potential as an ultra-sensitive detector~\cite{qmagnetNV}; even better performance can be achieved by applying more control pulses~\cite{Taylor,Naydenov}.

Here, we present a detailed experimental study of single NV center magnetometry with multi-pulse sensing sequences. We demonstrate that multi-pulse sequences greatly improve the sensitivity to oscillating magnetic fields. We find the optimal number of control pulses as function of field frequency and spin-echo decay time. These studies are performed both for fields with known phase and for fields with a phase that randomly fluctuates between measurement shots. We finally show that multi-pulse sequences can be used to achieve a high degree of frequency tunability and selectivity by exploiting their frequency-domain characteristics.

We use a single NV center in a nitrogen-rich Ib bulk diamond sample (Element Six) as our magnetometer. The setup is shown schematically in Fig. 1(a). We apply a static magnetic field oriented along the NV center's quantization axis $z$ that allows us to selectively address the $m_s = 0 \leftrightarrow m_s = -1$ spin transition. Within this subspace, the NV spin is equivalent to a spin-1/2. Using lithographically defined on-chip coplanar waveguides (CPWs)~\cite{fuchs} we achieve high fidelity ($\approx$99\%) control of the NV center electron spin state~\cite{bootstrap, ourscience}. The length of a single $\pi$-pulse used here is $t_p = 8$~ns. With the CPW low-frequency magnetic fields can be applied along the $z$-axis as well.

We consider a time-varying oscillating magnetic field oriented along the quantization axis $z$ of the probing spin $B_z(t) = b_z \sin(2 \pi f t +\phi)$. Components oscillating along the transverse axes will average to zero due to the rapid (2.5 GHz) precession of the NV spin. The working principle behind detecting a field of the form $B_z(t)$ is outlined in Fig. 1(b). An electron spin initialized in a superposition state precesses under the influence of the oscillating field. During each half-cycle the electron spin phase acquires $ \delta \Phi   = \int^{1/2f}_0 2 \pi \gamma B_z (t)dt = (2 \gamma b_z / f) \cos \phi $, with $\gamma = 28$ GHz/T for the electron spin of an NV center. Since the field oscillates the total phase after many cycles averages to zero. 
This can be prevented by using an $N$-pulse sequence with evenly spaced $\pi$-pulses 
\begin{equation}
\label{eq:CP}
[\tau/2 - \pi - \tau - \pi - \tau/2 ]^{N/2}.
\end{equation}
Note that this sequence has the same timing properties as the Carr-Purcell sequence (CP) \cite{CP} known from NMR. CP-like sequences have very recently been explored with NV centers in the context of dynamical decoupling from decoherence by a spin bath environment~\cite{ourscience, cory, Naydenov}. When the $\pi$-pulses of the sequence coincide with the nodes of $B_z(t)$, the electron accumulates a phase which increases with the length of the sequence. 
This can be understood by moving to the toggling reference frame of the electron spin from where $B_z(t) \rightarrow - B_z(t)$ after each $\pi$-pulse. Formally, $B_z(t)$ is multiplied by a time-domain filter function $Y(t)$ which changes sign each time a $\pi$-pulse is applied (Fig. 1(b)). After $N$ $\pi$-pulses (or half-cycles) the total phase of the electron spin state becomes $\Delta \Phi = N \delta \Phi$. 

\begin{figure}
\label{fig:f2}
\includegraphics[width=3.4in]{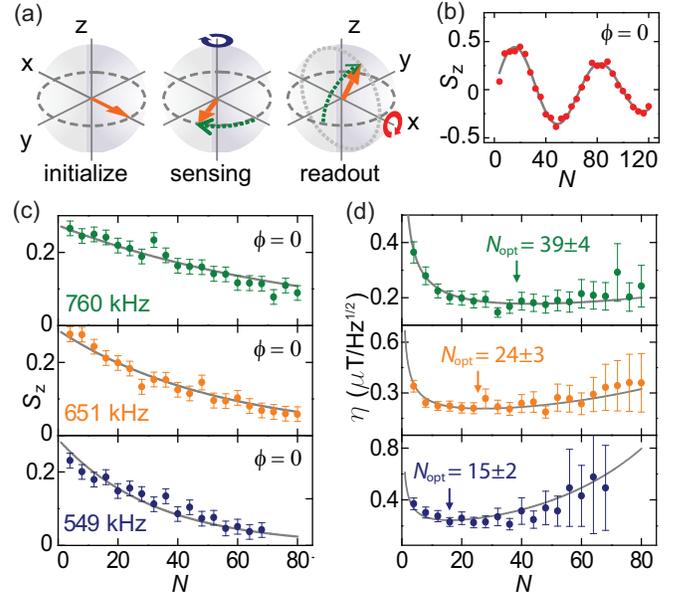}
\caption{Phase-locked magnetometry.  \textbf{(a)} Measurement scheme. A spin initialized along the $x$-axis will accumulate a phase during the sensing stage due to $B_z(t)$. A final rotation around the $x$-axis is applied to transform the phase into a population difference. \textbf{(b)} A 1 MHz ac field with constant amplitude is measured with increasing number of pulses $N$. Solid line is a fit to Eq. (\ref{eq:sz_ph_lo}), yielding $b_z=1.6$ $\mu$T. \textbf{(c)} Signal for three ac fields with different frequency. The amplitude is rescaled by $1/N$. Solid lines are fits to Eq. (\ref{eq:sz_ph_lo}) yielding $2N \gamma b_z / f \approx 0.2 \pi$ and $T_2 = (2.86\pm0.04) \mu s$
 \textbf{(d)} Sensitivities calculated from the data (points) and from the fits (solid lines) in (c).
}
\end{figure}

We first analyze the case when $B_z(t)$ always has the same phase relation with respect to the sequence of $\pi$-pulses ($\phi=0$)~\cite{degen,Taylor}. This corresponds to a situation where the phase of the field to be measured is under control of the experimenter (e.g. when one aims to detect spins that can be adiabatically inverted periodically). 
In this case, for a spin initialized along the $x$-axis, the read out can be performed by rotating the final state by $\pi/2$ around the same axis (see Fig. 2(a)). The resulting signal is then $S_z(b_z)= \frac{1}{2} \sin \left[ \left(2 N  \gamma b_z / f \right) \cos \phi \right]$ after normalization of the PL levels to $\left[ -\frac{1}{2}, \frac{1}{2} \right]$.

In Fig. 2(b) we monitor the evolution of the NV electron spin under application of an ac field with frequency $f=1$ MHz for increasing $N$. The ac field is phase-locked such that the $\pi$-pulses coincide with the nodes of $B_z(t)$. We observe oscillations which demonstrate spin precession in the applied ac field. The amplitude of the oscillations decays exponentially due to decoherence and pulse imperfections. We can limit the influence of pulse imperfections on the signal decay by using the XY4 sequence which is self-compensating for pulse errors \cite{XY4} and has the same timings as Eq. (\ref{eq:CP}). We find the maximum number of pulses that can be applied before pulse errors start to play a role to be $\sim 130$ pulses \cite{ourscience}. 

In our type Ib diamond sample the coherence time is limited by dipolar interactions with electron spins, which are located at the sites of substitutional nitrogen atoms in the diamond lattice~\cite{HansonandAwschalom}.
The one-pulse spin-echo signal decays as $\sim \exp[-(\tau/T_2)^3]$ where $T_2=(2.8\pm0.1)$ $\mu$s for the NV center used here~\cite{ourscience}. Increasing the number of pulses to $N$ reduces the signal by a factor $\sim \exp[-N(\tau/T_2)^3]$ for the CP sequence, which leads to the observed exponential decay in the signal in Fig. 2(b). For $B_z(t)$ oscillating with frequency $f=[2(\tau + t_p)]^{-1} \approx (2 \tau )^{-1}$ (assuming $t_p\ll\tau$) this gives a total signal of
\begin{equation}
\label{eq:sz_ph_lo}
S_z(b_z)= \frac{1}{2} \sin \left( \frac{2 N \gamma b_z }{f}  \cos \phi \right) e^{-N/(2f T_2)^3}.
\end{equation}
The optical detection of the spin population is limited by shot noise, which depends on the experimental parameters such as the number of photons collected per measurement shot $\varsigma$ and the contrast $C$ which combine to give the noise per measurement shot $\sigma _{ S_z } \approx 1/(C\sqrt{\varsigma})$~\cite{Taylor, Meriles}.
The sensitivity for detecting a field oscillating in phase ($\phi=0$) is given by combining the shot-noise limited minimum detectable field $b_{\mathrm{min}} = \sigma _{S_z} |\mathrm{d}b_z /\mathrm{d}  S_z |  \approx \sigma _{S_z} \frac{f}{ \gamma N}$ and the total integration time $T=N(\tau + t_p) = N/2f$ to give 
\begin{equation}
\label{eq:eta_ph_lo}
\eta\left(f,N\right) = b_{\mathrm{min}} \sqrt{T} = \frac{1}{ \gamma C}\sqrt{\frac{f}{2 \varsigma N}} e^{N/(2f T_2)^3}.
\end{equation}

It is instructive to consider two limiting cases. If $N \ll 2 f T_2$, decoherence is negligible and the use of a multi-pulse sequence improves the sensitivity by a factor $1/\sqrt{N}$. In the other extreme, where $N \gg 2 f T_2$, decoherence has a detrimental influence on the sensitivity: $\eta \propto e^{N/(2f T_2)^3}$. Thus, for a given frequency of $B_z (t)$ there exist an optimum number of pulses $N_{\mathrm{opt}}$. By minimizing $\eta$ we find $N_{\mathrm{opt}}=4T^3_2 f^3$.   
%
%

In Fig.~2(c) we demonstrate the detection of ac magnetic fields of three different frequencies, for increasing $N$. In order to keep the signal in the linear regime the amplitude of the field is rescaled as $N$ increases so that the total acquired phase after applying the sequence remains constant. For every measurement shot we also measure the zero-field signal to account for possible drifts in the setup. From the curves in Fig.~2(c) the sensitivity is calculated and depicted in Fig.~2(d), along with the predicted $N_{\mathrm{opt}}$ (calculated using the independently determined $T_2= (2.8\pm0.1)\mu$s). We observe that there indeed exists an optimum number of pulses for each frequency and find excellent agreement between the predicted $N_{\mathrm{opt}}$ and the data.

We now turn to the case where $B_z(t)$ has a phase which varies randomly between measurement shots, but remains constant during each individual measurement shot. The phase that a spin, initialized along $x$, acquires will be different for each measurement shot with zero average, as depicted in Fig. 3(a). Rotating the final state over the $x$ axis therefore yields zero signal when averaged over many measurements. However, as shown in Fig.~3(a),  the state can also be rotated around the (orthogonal) $y$-axis. This will give a signal, when averaged over the phase, of
\begin{eqnarray}
\label{eq:sz_nph_lo}
S_z (b_z)&=& \frac{1}{2} \left\langle \cos \left( \frac{2N \gamma b_z}{f} \cos \phi \right) \right\rangle_{\phi} e^{-N/(2f T_2)^3} \nonumber \\
&=& \frac{1}{2}J_0\left( \frac{2N \gamma b_z}{f} \right) e^{-N/(2f T_2)^3}
\end{eqnarray}
where $\left\langle \right\rangle_\phi$ denotes averaging over $\phi$ and $J_0$ is the zeroth-order Bessel function of the first kind. In the absence of any field $S_z (0) = \frac{1}{2}$. The data in Fig. 3(b) demonstrate that we can also detect the ac magnetic field if it has a random phase. 

\begin{figure}
\label{fig:f3}
\includegraphics[width=3.4in]{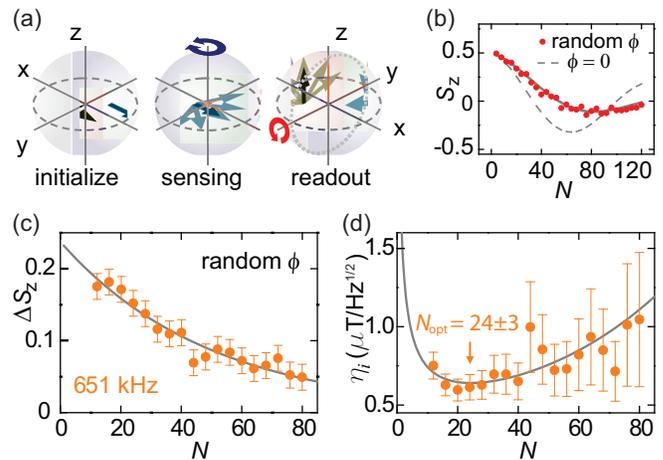}
\caption{Magnetometry for ac fields with random phase. \textbf{(a)} Measurement scheme. In each measurement shot the spin, initialized along the $x$-axis, accumulates a different phase during the sensing stage that depends on $\phi$. A signal can nonetheless be measured by applying the final $\pi /2$ rotation over the $y$-axis. \textbf{(b)} Measured signal for a field with constant amplitude and averaged over $\phi$ for increasing $N$. Solid line is a fit to Eq. (\ref{eq:sz_nph_lo}); the expected signal for $\phi=0$ (dashed) is included as a reference. \textbf{(c)} Signal intervals for a random-phase ac field of $651$~kHz. The amplitude $b_z$ is rescaled by $1/N$. Solid line is a fit to Eq. (\ref{eq:sz_nph_lo}) yielding $2N \gamma b_z/f \approx 0.47 \pi$ and $T_2 = (2.77\pm0.05) \mu s$. \textbf{(d)} Interval sensitivities calculated from the data (points) and from the fit (solid line) in (c).  
}
\end{figure}

A consequence of measuring the $x$-projection is that $|\mathrm{d}S_z /\mathrm{d}  b_z | \approx 2 N^2 \gamma^2 b_z/f^2$ will vanish as $b_z \rightarrow 0$, leading to a divergence in the differential sensitivity. We therefore turn to calculating the interval sensitivity \cite{Bauer} by extending the definition of the minimum detectable field per measurement shot to that measured for a given signal interval: $b_\mathrm{min} = \sigma_{S_z} b_z/\left| S_z(b_z)- S_z(0) \right|$. 
The interval sensitivity $\eta_i$ for measuring oscillating fields with random phase is given by 
\begin{equation}
\label{eq:non_ph_lo_sens}
\eta_i\left(f,N,b_z\right) = \frac{2 b_z \sigma_{S_z}}{\left| J_0\left(2N \gamma b_z/f\right)-1 \right|} \sqrt{\frac{N}{2 f}} e^{\frac{N}{(2fT_2)^3}}.
\end{equation} 

Analogously to the phase-locked experiment, we verify this expression experimentally in Fig. 3(c). The corresponding calculated interval sensitivities are depicted in Fig.~3(d). It shows a qualitatively similar picture as for the phase- locked case. Since we rescale $b_z$ by $1/N$, also here $N_{\mathrm{opt}}=4T^3_2 f^3$. Again, we observe that the theory gives an excellent description of the data.

Until now we discussed the situation where the sequence was tuned exactly in resonance with $B_z(t)$. In order to analyze what happens when the sequence is detuned with respect to $B_z(t)$ (so $\tau \neq \frac{1}{2f}$) we move to the frequency domain. For a sequence of evenly spaced pulses and $\phi=0$ the response of the magnetometer in the frequency domain is given by \cite{Taylor}

\begin{equation}
\label{eq:window_pl}
Y_N(f,\tau) = \frac{1-\sec(\pi f \tau)}{2 \pi \tau f} \sin(2 \pi N f \tau).
\end{equation}

Figure 4(a) depicts the signal detected as a function of frequency of $B_z(t)$ (with $\phi=0$) for $N=4$ to 80 for fixed $\tau$, mapping out the complete filter function Eq. (\ref{eq:window_pl}). Figure 4(b) shows line traces for three different $N$. With increasing $N$, the bandwidth of the response decreases by a factor $N$ while the peak signal increases $N$ times. Therefore, by tuning $\tau$ a single frequency component can be selected. This is useful for measuring the linewidth of $B_z(t)$ or for spectroscopic applications. The resolution is set by the full-width at half-maximum (FWHM) $\Delta f \approx  0.3 /(N \tau )$. 

\begin{figure}
\label{fig:f4}
\includegraphics[width=3.4in]{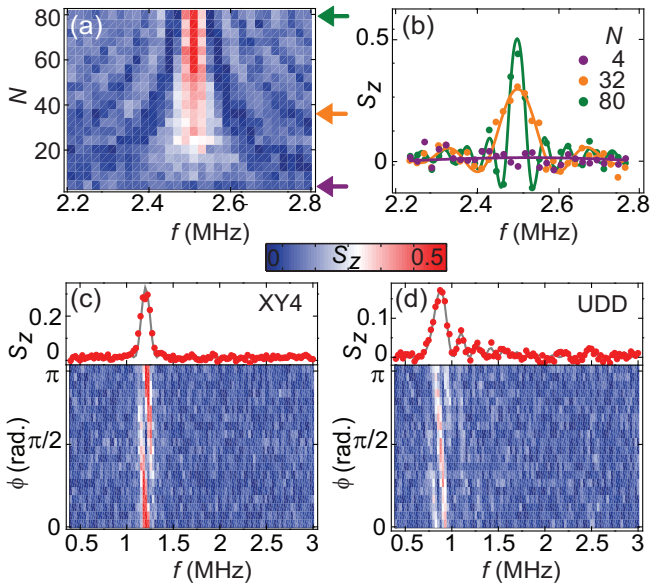}
\caption{Frequency domain analysis. \textbf{(a)} 
Measured filter response $Y_N (f, \tau )$ of the CP sequence vs. frequency for a field amplitude of $b_z \approx 0.8$  $\mu$T for $\phi=0$ and $\tau = 192$ ns. The scheme for measuring a phase-locked signal from Fig. 2(a) is used. A maximum signal is observed at $f=2(\tau + t_{\mathrm p})^{-1} = 2.5 $MHz. As $N$ is increased the bandwidth reduces and peak signal becomes higher. \textbf{(b)}  Line traces from (a). Solid lines are fits using Eq.~(\ref{eq:window_pl}). \textbf{(c)} Magnetometer response vs. frequency and phase using the CP-like XY4 sequence with $N=20$ pulses and 8.16 $\mu s$ integration time ($\tau = 400$ ns). The field amplitude is adjusted to $\sim 1.7$  $\mu$T. The scheme for measuring a random phase signal from Fig. 3(a) is used. \textbf{(d)} Signal for the same field and integration time measured in (c) but now using the 20-pulse aperiodic UDD sequence. The upper panels in (c) and (d) depict the signal averaged over the phase.
}
\end{figure}

We study the influence of the phase of $B_z(t)$ in more detail by performing a measurement at a fixed number of pulses for a range of initial phases. The results are depicted in Fig.~4(c). By averaging over the phase the response to fields with random phase is retrieved (see top panel of Fig. 4(c)). The filter function of sequences other than CP-like sequences can be investigated in a similar way. 

The Uhrig dynamical decoupling (UDD) sequence \cite{Uhrig} has been conjectured as a valuable tool in detecting randomly fluctuating fields \cite{Liam}. The $N$-pulse UDD sequence has pulses spaced irregularly according to $\tau_k = \sin ^2 [\pi k /(2N+2)]$ with $\tau_k$ the $k$-th pulse spacing. A characterization of the UDD sequence similar to Fig. 4(c) is presented in Fig. 4(d). UDD is seen to give a broader frequency response and reduced peak signal compared to a sequence with CP-like timings (Eq. (\ref{eq:CP})). The higher peak signal and reduced bandwidth of the latter is especially useful when gradients are used to achieve high spatial resolution. It will in general depend on the nature of $B_z(t)$ and the specific application for which sequence the best performance will be achieved.        

In conclusion, we have reported a detailed investigation of spin-based magnetometry with multi-pulse schemes. Our results show significantly enhanced performance both for ac fields with known and with unknown phase. These results pave the way towards unprecedented magnetic field sensitivity beyond the limit set by the spin-echo sequence. Note that the multi-pulse sequences also make the magnetometer insensitive to instabilities in the setup, such as drifts in the applied static magnetic field or in temperature~\cite{Acosta}. The insights gained here will help guide experimenters in tailoring the pulse sequence and number of control pulses to their specific application.

We gratefully acknowledge support from FOM, NWO, the EU SOLID and the DARPA QuEST program. Work at Ames Laboratory was supported by the Department of Energy --- Basic Energy Sciences under Contract No.~DE-AC02-07CH11358.

\end{document}